\documentclass[aps,twocolumn,showpacs]{revtex4}
\usepackage{graphicx}
\usepackage{dcolumn}
\usepackage{amsmath}
\usepackage{enumerate}

\begin{document}

\title{Uncertainties in nuclear transition matrix elements for neutrinoless 
$\beta \beta $ decay II: the heavy Majorana neutrino mass mechanism}

\author{P. K. Rath$^{1}$, R. Chandra$^{2,3}$, P. K. Raina$^{3,4}$, 
K. Chaturvedi$^{5}$, and J. G. Hirsch$^{6}$}

\affiliation{
$^{1}$Department of Physics, University of Lucknow, Lucknow-226007, India\\
$^{2}$Department of Applied Physics, Babasaheb Bhimrao Ambedkar University,
Lucknow-226025, India\\
$^{3}$Department of Physics and Meteorology, Indian Institute of Technology, 
Kharagpur-721302, India\\
$^{4}$Department of Physics, Indian Institute of Technology, Ropar,
Rupnagar - 140001, Punjab, India\\
$^{5}$Department of Physics, Bundelkhand University, Jhansi-284128, India\\
$^{6}$Instituto de Ciencias Nucleares, Universidad Nacional Aut\'{o}noma de
M\'{e}xico, 04510 M\'{e}xico, D.F., M\'{e}xico}
\date{\today}

\begin{abstract}
Employing four different parametrizations of the pairing plus multipolar
type of effective two-body interaction and three different parametrizations
of Jastrow-type of short range correlations, the  
uncertainties in the nuclear
transition matrix elements $M_{N}^{\left( 0\nu \right) }$ due to the exchange
of heavy Majorana neutrino for the $0^{+}\rightarrow 0^{+}$ transition of
neutrinoless double beta decay of $\ ^{94}$Zr, $^{96}$Zr, $^{98}$Mo, $^{100}$%
Mo, $^{104}$Ru, $^{110}$Pd, $^{128,130}$Te and $^{150}$Nd isotopes in the
PHFB model are estimated to be 
around 35\%. 
Excluding the nuclear transition matrix elements calculated with Miller-Spenser 
parametrization of Jastrow short range correlations, the uncertainties are 
found to be smaller than 20\%.
\end{abstract}

\pacs{21.60.-n, 23.40.-s, 23.40.Hc}

\maketitle

\section{INTRODUCTION}

In addition to establishing the Dirac or Majorana nature of neutrinos, the
observation of $\left( \beta \beta \right) _{0\nu }$ decay is a convenient
tool to test the lepton number conservation, possible hierarchies in the
neutrino mass spectrum, the 
origin of neutrino mass and CP violation in the
leptonic sector. Further, it can also ascertain the role of various gauge
models associated with all possible mechanisms, namely
the exchange of light neutrinos, heavy
neutrinos,
the right handed currents in the left-right symmetric
model (LRSM), the exchange of sleptons, neutralinos, squarks and gluinos in
the R$_{p}$-violating minimal super symmetric standard model, the
exchange of leptoquarks, existence of heavy sterile neutrinos,
compositeness, extradimensional scenarios and Majoron models, 
allowing the occurrence of $\left( \beta \beta \right) _{0\nu }$ decay. Stringent
limits on the associated  parameters have already been
extracted from the observed experimental limits on the half-life of $\left(
\beta ^{-}\beta ^{-}\right) _{0\nu }$ decay \cite{klap06} and presently, all
the experimental attempts are directed for its observation. The experimental
and theoretical studies devoted to $\left( \beta \beta \right) _{0\nu }$
decay over the past decades have been recently reviewed by Avignone 
\textit{\textit{et al.}} \cite{avig08} and references there in.

Presently, there is an increased interest to calculate reliable NTMEs for 
$\left( \beta ^{-}\beta ^{-}\right) _{0\nu }$
decay due to the exchange of heavy Majorana neutrinos,
in order to ascertain the
dominant mechanism contributing to it \cite{simk10,tell11}.  The lepton 
number violating $\left( \beta ^{-}\beta ^{-}\right) _{0\nu }$
decay has been studied by Vergados by taking a Lagrangian consisting of
left-handed as well as right-handed leptonic currents \cite{verg86}. In the
QRPA, the $\left( \beta ^{-}\beta ^{-}\right) _{0\nu }$ decay due to the
exchange of heavy Majorana neutrinos has been studied by Tomoda \cite{tomo91}%
. The decay rate of $\left( \beta ^{-}\beta ^{-}\right) _{0\nu }$ mode in
the LRSM has been derived by Doi and Kotani \cite{doi93}. Hirsch $et$ $al.
$ \cite{hirs96} have calculated all the required nuclear transition matrix
elements (NTMEs) in the QRPA and limits on the effective light neutrino mass 
$\left\langle m_{\nu }\right\rangle $%
, heavy neutrino mass $\left\langle M_{N}\right\rangle $, right handed heavy
neutrino $\left\langle M_{R}\right\rangle $, $\left\langle \lambda
\right\rangle $, $\left\langle \eta \right\rangle $ and mixing angle $tan\xi 
$ have been obtained. 
The heavy neutrino mechanism has also been studied in
the QRPA without \cite{pant94} and with pn-pairing \cite{pant96}. In the heavy
Majorana neutrino mass mechanism, \v{S}imkovic $et$ $al.$ \cite{simk92} have
studied the role of induced weak magnetism and pseudoscalar terms and it was
found that they are quite important in $^{48}$Ca nucleus. The importance of
the same induced currents in both light and heavy Majorana neutrino exchange
mechanism has also been studied using the pn-RQRPA \cite{simk01} as well as 
SRQRPA \cite{simk10}.

In spite of the remarkable 
success of the large scale shell model (LSSM) calculations of
Strassbourg-Madrid group \cite{stma}, there is a necessity of large 
configuration mixing
to reproduce the structural complexity of medium and heavy mass nuclei. On
the other hand, the QRPA and its extensions have emerged as 
successful models by including a large number of basis states and in
correlating the single-$\beta $ GT strengths and half-lives of ($\beta
^{-}\beta ^{-}$)$_{2\nu }$ decay in addition to explaining the observed
suppression of $M_{2\nu }$ \cite{voge86,civi87}. 
In the mass region $90\leq A\leq 150$, there is a subtle interplay of 
pairing and quadrupolar correlations and their effects 
on the NTMEs of $\left( \beta ^{-}\beta ^{-}\right) _{0\nu }$ decay 
have been studied in the interacting shell model (ISM) \cite{caur08,mene08},
deformed QRPA model \cite {pace04,alva04,yous09,fang10}, and
projected-Hartree-Fock-Bogoliubov (PHFB)\ model \cite{chat08,chan09}.
 
The possibility to constrain the values of the gauge parameters using the 
measured lower limits on the $\left( \beta ^{-}\beta ^{-}\right) _{0\nu }$ 
decay half-lives relies heavily on the model dependent NTMEs. Different 
predictions are obtained by employing different
nuclear models, and within a given model, varying the model space,
single particle energies (SPEs) and effective two-body interaction.
In addition, a number of issues regarding the structure of
NTMEs, namely the effect of pseudoscalar and
weak magnetism terms on the Fermi, Gamow-Teller and tensorial NTMEs \cite
{simk99,verg02}, the role of finite size of nucleons (FNS) as well as short
range correlations (SRC) vis-a-vis the radial evolution of NTMEs \cite
{simk08,simk09,caur08,rath09} and the value of the axial-vector coupling
constant $g_{A}$ are also the sources of uncertainties and 
remain to be investigated.

It was observed by Vogel \cite{voge00} that in case of well studied $^{76}$Ge, 
the
calculated decay rates $T{_{1/2}^{0\nu }}$ differ by a factor of 6-7 and
consequently, the uncertainty in the effective neutrino mass $\left\langle
m_{\nu }\right\rangle $ is about 2 to 3. Thus, the spread between the
calculated NTMEs can be used as the measure of the theoretical uncertainty.
In case the $\left( \beta \beta \right) _{0\nu }$ decay of different
nuclei will be observed, Bilenky and Grifols \cite{bile02} have suggested 
that the results of calculations of NTMEs of the 
$\left( \beta ^{-}\beta ^{-}\right)_{0\nu }$ decay can be checked by comparing 
the calculated ratios of the corresponding NTMEs-squared with the 
experimentally observed values.
  
Bahcall $et$ $al.$ \cite{bahc} and Avignone $et$ $al.$ \cite{avig06} have calculated averages of all the available NTMEs, and their standard deviation is taken as the
measure of theoretical uncertainty. On the other hand, Rodin \textit{\textit{%
et al. }}\cite{rodi03} have calculated nine NTMEs with three sets of basis
states and three realistic two-body effective interactions of charge
dependent Bonn, Argonne and Nijmen potentials in the QRPA as well as RQRPA
and estimated the theoretical uncertainties by making a statistical analysis.
It was noticed that the variances are substantially smaller
than the average
values and the results of QRPA, albeit slightly larger, are quite close to
the RQRPA values. Faessler and coworkers have further studied uncertainties
in NTMEs due to short range correlations using unitary correlation operator
method (UCOM) \cite{simk08} and
self-consistent coupled cluster method (CCM) \cite{simk09}.

The PHFB model has the advantage of treating the
pairing and deformation degrees of freedom on equal footing and projecting
out states with good angular momentum. However, the single $\beta $ decay rates 
and the distribution of GT strength, which require the structure of the 
intermediate odd $Z$-odd $N$ nuclei, can not be studied in the present version of
the PHFB model. In spite of this limitation, the PHFB model in
conjunction with pairing plus quadrupole-quadrupole (\textit{PQQ}) \cite
{bara68} has been successfully applied to reproduce the lowest yrast states,
electromagnetic properties of the parent and daughter nuclei, and the measured 
$\left( \beta ^{-}\beta^{-}\right) _{2\nu }$ decay rates \cite{chan05,sing07}. 
In the PHFB formalism, the existence of an inverse correlation between the 
quadrupole deformation and the size of NTMEs $M_{2\nu }$, $M^{\left( 0\nu \right) }$
and $M_{N}^{\left( 0\nu \right) }$ has been observed 
\cite{chat08,chan09}. Further, it has been noticed that the NTMEs are
usually large for a pair of spherical nuclei, almost constant for small
deformation, suppressed depending on the difference in the deformation $%
\Delta \beta _{2}$ of parent and daughter nuclei and having a well defined
maximum when $\Delta \beta _{2}=0$ \cite{chat08,chan09}.

In Ref. \cite{rath10}, a statistical analysis was performed for extracting
uncertainties in eight (twelve) NTMEs for $\left( \beta ^{-}\beta
^{-}\right) _{0\nu }$ decay due to the exchange of light Majorana neutrino, 
calculated in the PHFB model with four different
parameterizations of pairing plus multipolar type of effective two-body
interaction \cite{chan09} and two (three) different parametrization of
Jastrow type of SRC \cite{simk09}. In confirmation with the observation made
by \v{S}imkovic $et$ $al.$ \cite{simk09}, it was noticed that the Miller-Spenser
type of parametrization is a major source of uncertainty and its exclusion
reduces the uncertainties from 10\%--15\% to 4\%--14\%. Presently, the same
procedure has been adopted to estimate the theoretical uncertainties
associated with the NTMEs $M_{N}^{\left( 0\nu \right) }$ for $\left( \beta
^{-}\beta ^{-}\right) _{0\nu }$ decay due to the exchange of heavy Majorana
neutrino. In Sec. II, a brief discussion of the theoretical formalism is
presented. The results for different parameterizations of the two-body
interaction and SRC vis-a-vis radial evolution of NTMEs are discussed in Sec
III. In the same section, the averages as well as standard deviations are 
calculated for
estimating the theoretical uncertainties. Finally, the conclusions are given
in Sec. IV.

\section{THEORETICAL FORMALISM}

In the charged current weak processes, the current-current interaction under
the assumption of zero mass neutrinos leads to terms which, except for
vector and axial vector parts, are proportional to the lepton mass squared,
and hence negligible. However, it has been reported by \v{S}imkovic $et$ $al.$
\cite{simk99,verg02} that the contribution of the pseudoscalar term is
equivalent to a modification of the axial vector current due to PCAC and
greater than the vector current. The contributions of pseudoscalar and weak
magnetism terms in the mass mechanism can change $M^{(0\nu )}$ upto 30\% and the 
change in $M_{N}^{(0\nu )}$ is considerably larger. In the shell-model
\cite{caur08,horo10}, IBM \cite{bare09}
and GCM+PNAMP \cite{rodr10}, the contributions of these pseudoscalar and weak
magnetism terms to  $M^{(0\nu )}$ have been also investigated. However, it has 
been 
%%shown
reported
 by Suhonen and
Civitarese \cite{suho08} that these contributions are relatively small and
can be safely neglected. Therefore, the investigation of this issue is of
definite interest and is reported in the present work.

In the two nucleon mechanism\textit{,} the half-life $T_{1/2}^{0\nu }$ for
the $0^{+}\to 0^{+}$\ transition of $\left( \beta ^{-}\beta ^{-}\right)
_{0\nu }$ decay due to the exchange of heavy Majorana neutrino between
nucleons having finite size is given by \cite{tomo91,doi93}

\begin{equation}
\left[ T_{1/2}^{0\nu }\left( 0^{+}\to 0^{+}\right) \right] ^{-1}=\left(
\frac{m_{p}}{\left\langle M_{N}\right\rangle }\right) ^{2}G_{01}\left|
M_{N}^{\left( 0\nu \right) }\right| ^{2} ,
\end{equation}

\noindent where $m_{p}$ is the proton mass and
\begin{equation}
\left\langle M_{N}\right\rangle
^{-1}=\sum\nolimits_{i}U_{ei}^{2}m_{i}^{-1},\qquad \qquad m_{i}>1\text{ GeV},
\end{equation}
\noindent and in the closure approximation, the NTMEs $M_{N}^{\left( 0\nu \right) }$ is of the
form \cite{simk01,simk08,simk09}

\begin{equation}
M_{N}^{\left( 0\nu \right) }=-M_{Fh}+M_{GTh}+M_{Th} ,
\label{Mn}
\end{equation}
where
\begin{equation}
M_{\alpha }=\sum_{n,m}\left\langle 0_{F}^{+}\left\| O_{\alpha ,nm}\tau
_{n}^{+}\tau _{m}^{+}\right\| 0_{I}^{+}\right\rangle
\end{equation}
with
\begin{eqnarray}
O_{Fh} &=&H_{Fh}\left( r_{nm}\right)
\label{fh}
\\
O_{GTh}& =&\mathbf{\sigma }_{n}\cdot \mathbf{\sigma }_{m}H_{GTh}\left(
r_{nm}\right)  \\
O_{Th}& =&\left[3\left( \mathbf{\sigma }_{n}\cdot \widehat{\mathbf{r}}_{nm}\right)
\left( \mathbf{\sigma }_{m}\cdot \widehat{\mathbf{r}}_{nm}\right) -\mathbf{%
\sigma }_{n}\cdot \mathbf{\sigma }_{m} \right] H_{GTh}\left( r_{nm}\right)
\nonumber \\
\end{eqnarray}

The exchange of heavy Majorana neutrinos gives rise to short ranged neutrino potentials, which with the consideration of FNS are given by
\begin{eqnarray}
H_{\alpha h}(r_{nm}) &=&\frac{2R}{(m_{p}m_{e})\pi }\int f_{\alpha h}\left(
qr_{nm}\right) h_{\alpha }(q)q^{2}dq
\end{eqnarray}
where $f_{\alpha h}\left( qr_{nm}\right) =j_{0}\left( qr_{nm}\right) $  
for $\alpha=F $ as well as $GT$ and
$f_{Th}\left( qr_{nm}\right) =j_{2}\left( qr_{nm}\right)$.
%where $f_{F,GT h}\left( qr_{nm}\right) =j_{0}\left( qr_{nm}\right) $ and $%
%f_{T h}\left( qr_{nm}\right) =j_{2}\left( qr_{nm}\right) $ . 

Further, the $h_{F}(q)$, $h_{GT}(q)$ and 
$h_{T}(q)$
are written as
\begin{widetext}
\begin{eqnarray}
h_{F}(q) &=&\left( \frac{g_{V}}{g_{A}}\right) ^{2}\left( \frac{\Lambda
_{V}^{2}}{q^{2}+\Lambda _{V}^{2}}\right) ^{4} \label{hF}\\
h_{GT}(q) &=&\frac{g_{A}^{2}(q^{2})}{g_{A}^{2}}
\left[ 1-\frac{2}{3}\frac{g_{P}(q^{2})q^{2}}{g_{A}(q^{2})2m_{p}}
%%\left[ 1-\frac{2}{3}\frac{g_{P}(q^{2})q^{2}}{2m_{p}}
+\frac{1}{3}\frac{g_{P}^{2}(q^{2})q^{4}}{g_{A}^{2}(q^{2})4m_{p}^{2}}\right] 
+\frac{2}{3}\frac{g_{M}^{2}(q^{2})q^{2}}{g_{A}^{2}4m_{p}^{2}}  \nonumber \\
&\approx &\left( \frac{\Lambda _{A}^{2}}{q^{2}+\Lambda _{A}^{2}}\right)
^{4}\left[ 1-\frac{2}{3}\frac{q^{2}}{\left( q^{2}+m_{\pi }^{2}\right) }+%
\frac{1}{3}\frac{q^{4}}{\left( q^{2}+m_{\pi }^{2}\right) ^{2}}\right]
+\left( \frac{g_{V}}{g_{A}}\right) ^{2}\frac{\kappa ^{2}q^{2}}{6m_{p}^{2}}%
\left( \frac{\Lambda _{V}^{2}}{q^{2}+\Lambda _{V}^{2}}\right) ^{4} \label{hGT}\\
h_{T}(q) &=&\frac{g_{A}^{2}(q^{2})}{g_{A}^{2}}
\left[ \frac{2}{3}\frac{g_{P}(q^{2})q^{2}}{g_{A}(q^{2})2m_{p}}
%\left[ \frac{2}{3}\frac{g_{P}(q^{2})q^{2}}{2m_{p}}
-\frac{1}{3}\frac{g_{P}^{2}(q^{2})q^{4}}{g_{A}^{2}(q^{2})4m_{p}^{2}}\right] 
+\frac{1}{3}\frac{g_{M}^{2}(q^{2})q^{2}}{g_{A}^{2}4m_{p}^{2}}  \nonumber \\
&\approx &\left( \frac{\Lambda _{A}^{2}}{q^{2}+\Lambda _{A}^{2}}\right)
^{4}\left[ \frac{2}{3}\frac{q^{2}}{\left( q^{2}+m_{\pi }^{2}\right) }-\frac{1%
}{3}\frac{q^{4}}{\left( q^{2}+m_{\pi }^{2}\right) ^{2}}\right] +\left( \frac{%
g_{V}}{g_{A}}\right) ^{2}\frac{\kappa ^{2}q^{2}}{12m_{p}^{2}}\left( \frac{%
\Lambda _{V}^{2}}{q^{2}+\Lambda _{V}^{2}}\right) ^{4} \label{hT}
\end{eqnarray}
\end{widetext}
where the form factors are given by
\begin{eqnarray}
%g_{V}(q^{2}) &=&g_{V}\left( \dfrac{\Lambda _{V}^{2}}{q^{2}+\Lambda _{V}^{2}}%
%\right) ^{2}  \nonumber \\
g_{A}(q^{2}) &=&g_{A}\left( \dfrac{\Lambda _{A}^{2}}{q^{2}+\Lambda _{A}^{2}}%
\right) ^{2}  \nonumber \\
%%g_{M}(q^{2}) &=&\kappa g_{V}\left( q^{2}\right)  \nonumber \\
g_{M}(q^{2}) &=&\kappa g_{V}\left( \dfrac{\Lambda _{V}^{2}}{q^{2}+\Lambda _{V}^{2}}%
\right) ^{2}  \nonumber \\
g_{P}(q^{2}) &=&\dfrac{2m_{p}g_{A}(q^{2})}{\left( q^{2}+m_{\pi }^{2}\right) }%
\left( \dfrac{\Lambda _{A}^{2}-m_{\pi }^{2}}{\Lambda _{A}^{2}}\right)
\end{eqnarray}
with $g_{V}=1.0$, $g_{A}=1.254$, $\kappa =\mu _{p}-\mu _{n}=3.70$, $\Lambda
_{V}=0.850$ GeV, $\Lambda _{A}=1.086$ GeV
and $m_{\pi}$ is the pion mass.

Substituting Eq. (\ref{fh})--Eq. (\ref{hT}) in Eq. (\ref{Mn}), there is one term, 
associated with $h_F$, Eq. (\ref{hF}), 
contributing to $M_{Fh}$, while $M_{GTh}$ has four terms, denoted by $M_{GT-AA}$,   
$M_{GT-AP}$, $M_{GT-PP}$  and $M_{GT-MM}$, which correspond to the four terms in  
$h_{GT}$, Eq. (\ref{hGT}).
The tensor contribution, $M_{Th}$, has three terms, denoted by $M_{T-AP}$,
$M_{T-PP}$  and $M_{T-MM}$, which correspond to the three terms in  $h_{T}$, 
Eq. (\ref{hT}).
Their contributions to the total nuclear matrix element are discussed in Sec. III.

The short range correlations (SRC) arise mainly from the repulsive
nucleon-nucleon potential due to the exchange of $\rho $ and $\omega $
mesons and have been incorporated by using effective transition operator
\cite{wu85}, the exchange of $\omega $-meson \cite{jghi95}, UCOM \cite
{kort07,simk08} and the self-consistent CCM \cite{simk09}.
The SRC can also be incorporated phenomenologically by Jastrow type of
correlations with Miller-Spenser parametrization \cite{mill76}.
Further, it has been shown in the self-consistent CMM \cite{simk09} that the
SRC effects of Argonne and CD-Bonn two nucleon potentials are weak and it is
possible to parametrize them by Jastrow type of correlations within a few
percent accuracy. Explicitly,

\begin{equation}
f(r)=1-ce^{-ar^{2}}(1-br^{2})
\end{equation}
where $a=1.1$, $1.59$ and $1.52$ $fm^{-2}$, $b=0.68$, $1.45$ and $1.88$ $%
fm^{-2}$ and $c=1.0$, $0.92$ and $0.46$ for Miller-Spencer parametrization,
CD-Bonn and Argonne V18 NN potentials, respectively. 
%%Presently,
In this work
 the NTMEs $M_{N}^{\left( 0\nu \right) }$ are calculated in the PHFB model 
for the above
mentioned three sets of parameters for the SRC, denoted as SRC1, SRC2 and
SRC3, respectively.

\begin{figure}[htbp]
\includegraphics [scale=0.7,angle=270]{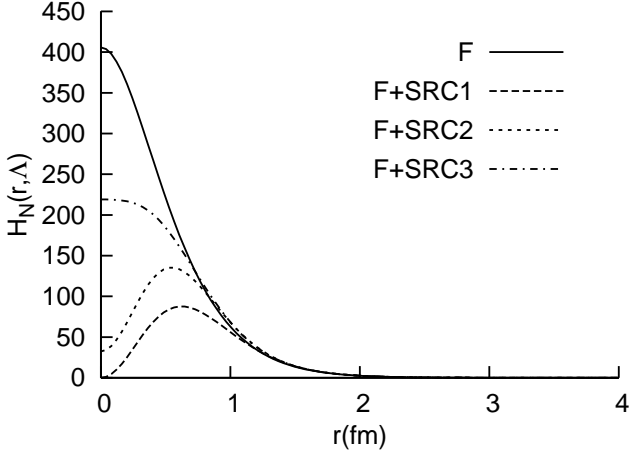}
\caption{Radial dependence of $H_{N}\left( r,\Lambda \right)$= 
$H_{Fh}\left( r,\Lambda \right)f(r)$
for the three different parameterizations of the SRC. In the 
case of FNS, $f(r)=1$.}
\label{fig1}
\end{figure}

In Fig.1, we plot the neutrino potential 
$H_{N}\left( r,\Lambda \right)$=
$H_{Fh}\left( r,\Lambda \right)f(r)$
with the three different parametrizations of SRC.   
It is noticed, that the potentials 
due to FNS and FNS+SRC3 are peaked at the origin where as the peaks due to 
FNS+SRC1 
and FNS+SRC3 are at $r\approx 0.6$ fm and $r\approx 0.5$ fm, respectively. The 
shapes of these functions have definite influence on the radial evolution of 
NTMEs  $M_{N}^{(0\nu )}$ for  $\left( \beta ^{-}\beta ^{-}\right)_{0\nu }$ 
decay due to the exchange of heavy Majorana neutrino as discussed in Sec. III.

The calculation of $M_{N}^{\left( 0\nu \right) }$ in the PHFB model has been
discussed in our earlier work \cite{chat08,rath10} and one obtains the
following expression for NTMEs $M_{\alpha }^{\left( 0\nu \right) }$ of $%
\left( \beta ^{-}\beta ^{-}\right) _{0\nu }$ decay \cite{rath10} 
\begin{widetext}
\begin{eqnarray}
M_{\alpha }^{\left( 0\nu \right) } &=&\left[ n^{Ji=0}n^{J_{f}=0}\right]
^{-1/2}\int\limits_{0}^{\pi }n_{(Z,N),(Z+2,N-2)}(\theta )\sum\limits_{\alpha
\beta \gamma \delta }\left( \alpha \beta \left| O_{\alpha }\right| \gamma
\delta \right)  \nonumber \\
&&\times \sum\limits_{\varepsilon \eta }\frac{\left( f_{Z+2,N-2}^{(\pi
)*}\right) _{\varepsilon \beta }}{\left[ \left( 1+F_{Z,N}^{(\pi )}(\theta
)f_{Z+2,N-2}^{(\pi )*}\right) \right] _{\varepsilon \alpha }}\frac{%
\left( F_{Z,N}^{(\nu )*}\right) _{\eta \delta }}{\left[ \left(
1+F_{Z,N}^{(\nu )}(\theta )f_{Z+2,N-2}^{(\nu )*}\right) \right] _{\gamma
\eta }}sin\theta d\theta  \label{ntm}
\end{eqnarray}
\end{widetext}
and the expressions for calculating $n^{J}$, $n_{(Z,N),(Z+2,N-2)}{(\theta )}$%
, $f_{Z,N}$ and $F_{Z,N}(\theta )$ are given in Refs. \cite{chat08,rath10}.

The calculation of matrices $f_{Z,N}$ and $F_{Z,N}(\theta )$ requires the
amplitudes $(u_{im},v_{im})$ and expansion coefficients $C_{ij,m}$, which
specify the axially symmetric HFB intrinsic state ${|\Phi _{0}\rangle }$
with $K=0$. Presently, they are obtained by carrying out the HFB
calculations through the minimization of the expectation value of the
effective Hamiltonian given by \cite{chan09} 
\begin{equation}
H=H_{sp}+V(P)+V(QQ)+V(HH)
\end{equation}
where $H_{sp}$, $V(P)$, $V(QQ)$ and $V(HH)$ denote the single particle
Hamiltonian, the pairing, quadrupole-quadrupole and
hexadecapole-hexadecapole part of the effective two-body interaction,
respectively. The $HH$ part of the effective interaction $V(HH)$ is written
as \cite{chan09} 
\begin{widetext}
\begin{equation}
V(HH)=-\left( \frac{\chi _{4}}{2}\right) \sum\limits_{\alpha \beta \gamma
\delta }\sum\limits_{\nu }(-1)^{\nu }\langle \alpha |r^{4}Y_{4,\nu }(\theta
,\phi )|\gamma \rangle \langle \beta |r^{4}Y_{4,-\nu }(\theta ,\phi )|\delta
\rangle \ a_{\alpha }^{\dagger }a_{\beta }^{\dagger }\ a_{\delta }\
a_{\gamma }
\end{equation}
\end{widetext}
\noindent with $\chi _{4}=0.2442$ $\chi _{2}A^{-2/3}b^{-4}$ for $T=1$, and 
twice of this value for $T=0$ case, following Bohr and Mottelson \cite{bohr98}.

In Refs. \cite{chan05,sing07,chat08}, the strengths of the like particle
components $\chi _{pp}$ and $\chi _{nn}$ of the $QQ$ interaction were
kept fixed. The strength of proton-neutron (\textit{pn}) component $\chi
_{pn}$ was varied so as to reproduce the excitation energy of the $\ $2$^{+}$
state \ $E_{2^{+}}$ for the considered nuclei, namely $^{94,96}$Zr, 
$^{94,96,98,100}$Mo, $^{98,100,104}$Ru, $^{104,110}$Pd, $^{110}$Cd, 
$^{128,130}$Te, $^{128,130}$Xe, $^{150}$Nd and $^{150}$Sm as closely possible 
to the experimental
values. This is denoted as $PQQ1$ parametrization. Alternatively, one
can employ a different parametrization of the $\chi _{2pn}$, namely 
$PQQ2$ by taking $\chi _{2pp}=\chi _{2nn}=\chi _{2pn}/2$ and the excitation
energy $E_{2^{+}}$ can be reproduced by varying the $\chi _{2pp}$. Adding
the $HH$ part of the two-body interaction to $PQQ1$ and
$PQQ2$ and by repeating the calculations, two more parameterizations of the
effective two-body interactions, namely $PQQHH1$ and $PQQHH2$
were obtained \cite{rath10}.

The four different parameterizations of the effective pairing plus multipolar
correlations provide us four different sets of wave functions. With three
different parameterizations of Jastrow type of SRC and four sets of wave
functions, sets of twelve NTMEs $M_{N}^{\left( 0\nu \right) }$ are
calculated for estimating the associated
uncertainties in the present work. The uncertainties associated with the
NTMEs $M_{N}^{(0\nu )}$ for $\left( \beta ^{-}\beta ^{-}\right) _{0\nu }$
decay are estimated statistically by calculating the mean and the standard
deviation defined by 
\begin{equation}
\overline{M}_{N}^{(0\nu )}=\frac{\sum_{i=1}^{k}M_{N}^{(0\nu )}(i)}{N}
\label{avrg}
\end{equation}
and 
\begin{equation}
\Delta \overline{M}_{N}^{(0\nu )}=\frac{1}{\sqrt{N-1}}\left[
\sum_{i=1}^{N}\left( \overline{M}_{N}^{(0\nu )}-M_{N}^{(0\nu )}(i)\right)
^{2}\right] ^{1/2}  \label{stdv}
\end{equation}

\section{RESULTS AND DISCUSSIONS}

The model space, SPE's, parameters of $PQQ$ type of effective two-body
interactions and the method to fix them have already been given in Refs. 
\cite{chan05,sing07,chat08}. It turns out that with $PQQ1$ and 
$PQQ2$ parameterizations, the experimental excitation energies of the 2%
$^{+}$ state \ $E_{2^{+}}$ \cite{saka84} can be reproduced within about 2\%
accuracy. The electromagnetic properties, namely reduced $B(E2$:$0^{+}\to
2^{+})$ transition probabilities, deformation parameters $\beta _{2}$,
static quadrupole moments $Q(2^{+})$ and gyromagnetic factors $g(2^{+})$ are
in overall agreement with the experimental data \cite{ragh89,rama01}.

\begin{table}[htbp]
\caption{Calculated NTMEs $M_{N}^{\left( 0\nu \right) }$ in the PHFB model
with four different parameterization of effective two-body interaction and
three different parameterizations of Jastrow type of SRC for the $\left( \beta
^{-}\beta ^{-}\right) _{0\nu }$ \ decay of $^{94,96}$Zr, $^{98,100}$Mo, $%
^{104}$Ru, $^{110}$Pd, $^{128,130}$Te and $^{150}$Nd isotopes due to the
exchange of heavy Majorana neutrino exchange. 
(a), (b), (c) and (d) denote $PQQ1$, $PQQHH1$, $PQQ2$ and $PQQHH2$ 
parameterizations, respectively.
See the footnote in page 3 of Ref. \cite{rath10} for further details.}
\label{tab1}
\begin{tabular}{rrrccccccc}
\hline\hline
{\small Nuclei} &~~~~~~  &  & {\small F} &  & \multicolumn{5}{c}{\small F+S} \\
\cline{6-10}
&  &  &  &  & {\small SRC1} &  & {\small SRC2} &  & {\small SRC3} \\ \hline
$^{94}${\small Zr} & {\small (a)} &  & \multicolumn{1}{r}{\small 236.9498} &
\multicolumn{1}{r}{} & \multicolumn{1}{r}{\small 77.5817} &
\multicolumn{1}{r}{} & \multicolumn{1}{r}{\small 138.2606} &
\multicolumn{1}{r}{} & \multicolumn{1}{r}{\small 191.3897} \\
& {\small (b)} &  & \multicolumn{1}{r}{\small 220.3794} & \multicolumn{1}{r}{
} & \multicolumn{1}{r}{\small 72.4285} & \multicolumn{1}{r}{} &
\multicolumn{1}{r}{\small 128.7496} & \multicolumn{1}{r}{} &
\multicolumn{1}{r}{\small 178.0783} \\
& {\small (c)} &  & \multicolumn{1}{r}{\small 205.8370} & \multicolumn{1}{r}{
} & \multicolumn{1}{r}{\small 72.9303} & \multicolumn{1}{r}{} &
\multicolumn{1}{r}{\small 124.3248} & \multicolumn{1}{r}{} &
\multicolumn{1}{r}{\small 168.5705} \\
& {\small (d)} &  & \multicolumn{1}{r}{\small 211.0437} & \multicolumn{1}{r}{
} & \multicolumn{1}{r}{\small 68.9323} & \multicolumn{1}{r}{} &
\multicolumn{1}{r}{\small 122.9710} & \multicolumn{1}{r}{} &
\multicolumn{1}{r}{\small 170.3572} \\
%&  &  & \multicolumn{1}{r}{} & \multicolumn{1}{r}{} & \multicolumn{1}{r}{} &
%\multicolumn{1}{r}{} & \multicolumn{1}{r}{} & \multicolumn{1}{r}{} &
%\multicolumn{1}{r}{} \\
$^{96}${\small Zr} & {\small (a)} &  & \multicolumn{1}{r}{\small 177.7479} &
\multicolumn{1}{r}{} & \multicolumn{1}{r}{\small 56.4909} &
\multicolumn{1}{r}{} & \multicolumn{1}{r}{\small 102.4434} &
\multicolumn{1}{r}{} & \multicolumn{1}{r}{\small 142.8831} \\
& {\small (b)} &  & \multicolumn{1}{r}{\small 185.5251} & \multicolumn{1}{r}{
} & \multicolumn{1}{r}{\small 59.5338} & \multicolumn{1}{r}{} &
\multicolumn{1}{r}{\small 107.2877} & \multicolumn{1}{r}{} &
\multicolumn{1}{r}{\small 149.3117} \\
& {\small (c)} &  & \multicolumn{1}{r}{\small 170.8199} & \multicolumn{1}{r}{
} & \multicolumn{1}{r}{\small 54.2382} & \multicolumn{1}{r}{} &
\multicolumn{1}{r}{\small 98.4051} & \multicolumn{1}{r}{} &
\multicolumn{1}{r}{\small 137.2870} \\
& {\small (d)} &  & \multicolumn{1}{r}{\small 175.4730} & \multicolumn{1}{r}{
} & \multicolumn{1}{r}{\small 56.0746} & \multicolumn{1}{r}{} &
\multicolumn{1}{r}{\small 101.2963} & \multicolumn{1}{r}{} &
\multicolumn{1}{r}{\small 141.1240} \\
%&  &  & \multicolumn{1}{r}{} & \multicolumn{1}{r}{} & \multicolumn{1}{r}{} &
%\multicolumn{1}{r}{} & \multicolumn{1}{r}{} & \multicolumn{1}{r}{} &
%\multicolumn{1}{r}{} \\
$^{98}${\small Mo} & {\small (a)} &  & \multicolumn{1}{r}{\small 355.1915} &
\multicolumn{1}{r}{} & \multicolumn{1}{r}{\small 117.0804} &
\multicolumn{1}{r}{} & \multicolumn{1}{r}{\small 208.2494} &
\multicolumn{1}{r}{} & \multicolumn{1}{r}{\small 287.5615} \\
& {\small (b)} &  & \multicolumn{1}{r}{\small 346.1118} & \multicolumn{1}{r}{
} & \multicolumn{1}{r}{\small 116.4967} & \multicolumn{1}{r}{} &
\multicolumn{1}{r}{\small 204.5667} & \multicolumn{1}{r}{} &
\multicolumn{1}{r}{\small 281.0515} \\
& {\small (c)} &  & \multicolumn{1}{r}{\small 358.5109} & \multicolumn{1}{r}{
} & \multicolumn{1}{r}{\small 118.0563} & \multicolumn{1}{r}{} &
\multicolumn{1}{r}{\small 210.1150} & \multicolumn{1}{r}{} &
\multicolumn{1}{r}{\small 290.2080} \\
& {\small (d)} &  & \multicolumn{1}{r}{\small 343.4160} & \multicolumn{1}{r}{
} & \multicolumn{1}{r}{\small 115.2077} & \multicolumn{1}{r}{} &
\multicolumn{1}{r}{\small 202.6977} & \multicolumn{1}{r}{} &
\multicolumn{1}{r}{\small 278.7158} \\
%&  &  & \multicolumn{1}{r}{} & \multicolumn{1}{r}{} & \multicolumn{1}{r}{} &
%\multicolumn{1}{r}{} & \multicolumn{1}{r}{} & \multicolumn{1}{r}{} &
%\multicolumn{1}{r}{} \\
$^{100}${\small Mo} & {\small (a)} &  & \multicolumn{1}{r}{\small 365.8004}
& \multicolumn{1}{r}{} & \multicolumn{1}{r}{\small 122.2000} &
\multicolumn{1}{r}{} & \multicolumn{1}{r}{\small 215.8882} &
\multicolumn{1}{r}{} & \multicolumn{1}{r}{\small 296.9869} \\
& {\small (b)} &  & \multicolumn{1}{r}{\small 361.9877} & \multicolumn{1}{r}{
} & \multicolumn{1}{r}{\small 122.6611} & \multicolumn{1}{r}{} &
\multicolumn{1}{r}{\small 214.7455} & \multicolumn{1}{r}{} &
\multicolumn{1}{r}{\small 294.4297} \\
& {\small (c)} &  & \multicolumn{1}{r}{\small 368.4056} & \multicolumn{1}{r}{
} & \multicolumn{1}{r}{\small 123.2364} & \multicolumn{1}{r}{} &
\multicolumn{1}{r}{\small 217.5391} & \multicolumn{1}{r}{} &
\multicolumn{1}{r}{\small 299.1598} \\
& {\small (d)} &  & \multicolumn{1}{r}{\small 328.9795} & \multicolumn{1}{r}{
} & \multicolumn{1}{r}{\small 111.4464} & \multicolumn{1}{r}{} &
\multicolumn{1}{r}{\small 195.1601} & \multicolumn{1}{r}{} &
\multicolumn{1}{r}{\small 267.5869} \\
%&  &  & \multicolumn{1}{r}{} & \multicolumn{1}{r}{} & \multicolumn{1}{r}{} &
%\multicolumn{1}{r}{} & \multicolumn{1}{r}{} & \multicolumn{1}{r}{} &
%\multicolumn{1}{r}{} \\
$^{104}${\small Ru} & {\small (a)} &  & \multicolumn{1}{r}{\small 274.0700}
& \multicolumn{1}{r}{} & \multicolumn{1}{r}{\small 89.7666} &
\multicolumn{1}{r}{} & \multicolumn{1}{r}{\small 160.7925} &
\multicolumn{1}{r}{} & \multicolumn{1}{r}{\small 222.1151} \\
& {\small (b)} &  & \multicolumn{1}{r}{\small 264.9015} & \multicolumn{1}{r}{
} & \multicolumn{1}{r}{\small 88.1515} & \multicolumn{1}{r}{} &
\multicolumn{1}{r}{\small 156.2893} & \multicolumn{1}{r}{} &
\multicolumn{1}{r}{\small 215.1076} \\
& {\small (c)} &  & \multicolumn{1}{r}{\small 258.2796} & \multicolumn{1}{r}{
} & \multicolumn{1}{r}{\small 84.6746} & \multicolumn{1}{r}{} &
\multicolumn{1}{r}{\small 151.6002} & \multicolumn{1}{r}{} &
\multicolumn{1}{r}{\small 209.3600} \\
& {\small (d)} &  & \multicolumn{1}{r}{\small 247.0603} & \multicolumn{1}{r}{
} & \multicolumn{1}{r}{\small 82.3208} & \multicolumn{1}{r}{} &
\multicolumn{1}{r}{\small 145.8435} & \multicolumn{1}{r}{} &
\multicolumn{1}{r}{\small 200.6645} \\
%&  &  & \multicolumn{1}{r}{} & \multicolumn{1}{r}{} & \multicolumn{1}{r}{} &
%\multicolumn{1}{r}{} & \multicolumn{1}{r}{} & \multicolumn{1}{r}{} &
%\multicolumn{1}{r}{} \\
$^{110}${\small Pd} & {\small (a)} &  & \multicolumn{1}{r}{\small 424.6601}
& \multicolumn{1}{r}{} & \multicolumn{1}{r}{\small 140.3359} &
\multicolumn{1}{r}{} & \multicolumn{1}{r}{\small 249.6835} &
\multicolumn{1}{r}{} & \multicolumn{1}{r}{\small 344.3187} \\
& {\small (b)} &  & \multicolumn{1}{r}{\small 379.9404} & \multicolumn{1}{r}{
} & \multicolumn{1}{r}{\small 127.4915} & \multicolumn{1}{r}{} &
\multicolumn{1}{r}{\small 224.6563} & \multicolumn{1}{r}{} &
\multicolumn{1}{r}{\small 308.6907} \\
& {\small (c)} &  & \multicolumn{1}{r}{\small 407.2163} & \multicolumn{1}{r}{
} & \multicolumn{1}{r}{\small 134.6824} & \multicolumn{1}{r}{} &
\multicolumn{1}{r}{\small 239.4733} & \multicolumn{1}{r}{} &
\multicolumn{1}{r}{\small 330.1888} \\
& {\small (d)} &  & \multicolumn{1}{r}{\small 390.3539} & \multicolumn{1}{r}{
} & \multicolumn{1}{r}{\small 130.6314} & \multicolumn{1}{r}{} &
\multicolumn{1}{r}{\small 230.5392} & \multicolumn{1}{r}{} &
\multicolumn{1}{r}{\small 316.9996} \\
%&  &  & \multicolumn{1}{r}{} & \multicolumn{1}{r}{} & \multicolumn{1}{r}{} &
%\multicolumn{1}{r}{} & \multicolumn{1}{r}{} & \multicolumn{1}{r}{} &
%\multicolumn{1}{r}{} \\
$^{128}${\small Te} & {\small (a)} &  & \multicolumn{1}{r}{\small 190.5325}
& \multicolumn{1}{r}{} & \multicolumn{1}{r}{\small 62.4373} &
\multicolumn{1}{r}{} & \multicolumn{1}{r}{\small 111.5143} &
\multicolumn{1}{r}{} & \multicolumn{1}{r}{\small 154.1796} \\
& {\small (b)} &  & \multicolumn{1}{r}{\small 231.8024} & \multicolumn{1}{r}{
} & \multicolumn{1}{r}{\small 77.4559} & \multicolumn{1}{r}{} &
\multicolumn{1}{r}{\small 136.7936} & \multicolumn{1}{r}{} &
\multicolumn{1}{r}{\small 188.1893} \\
& {\small (c)} &  & \multicolumn{1}{r}{\small 220.7156} & \multicolumn{1}{r}{
} & \multicolumn{1}{r}{\small 73.5158} & \multicolumn{1}{r}{} &
\multicolumn{1}{r}{\small 130.0810} & \multicolumn{1}{r}{} &
\multicolumn{1}{r}{\small 179.0960} \\
& {\small (d)} &  & \multicolumn{1}{r}{\small 235.4814} & \multicolumn{1}{r}{
} & \multicolumn{1}{r}{\small 78.6367} & \multicolumn{1}{r}{} &
\multicolumn{1}{r}{\small 138.9052} & \multicolumn{1}{r}{} &
\multicolumn{1}{r}{\small 191.1366} \\
%&  &  & \multicolumn{1}{r}{} & \multicolumn{1}{r}{} & \multicolumn{1}{r}{} &
%\multicolumn{1}{r}{} & \multicolumn{1}{r}{} & \multicolumn{1}{r}{} &
%\multicolumn{1}{r}{} \\
$^{130}${\small Te} & {\small (a)} &  & \multicolumn{1}{r}{\small 236.0701}
& \multicolumn{1}{r}{} & \multicolumn{1}{r}{\small 81.5493} &
\multicolumn{1}{r}{} & \multicolumn{1}{r}{\small 141.3447} &
\multicolumn{1}{r}{} & \multicolumn{1}{r}{\small 192.7610} \\
& {\small (b)} &  & \multicolumn{1}{r}{\small 231.5921} & \multicolumn{1}{r}{
} & \multicolumn{1}{r}{\small 79.3844} & \multicolumn{1}{r}{} &
\multicolumn{1}{r}{\small 138.1901} & \multicolumn{1}{r}{} &
\multicolumn{1}{r}{\small 188.8492} \\
& {\small (c)} &  & \multicolumn{1}{r}{\small 233.0024} & \multicolumn{1}{r}{
} & \multicolumn{1}{r}{\small 80.4020} & \multicolumn{1}{r}{} &
\multicolumn{1}{r}{\small 139.4400} & \multicolumn{1}{r}{} &
\multicolumn{1}{r}{\small 190.2194} \\
& {\small (d)} &  & \multicolumn{1}{r}{\small 230.5282} & \multicolumn{1}{r}{
} & \multicolumn{1}{r}{\small 78.9888} & \multicolumn{1}{r}{} &
\multicolumn{1}{r}{\small 137.5307} & \multicolumn{1}{r}{} &
\multicolumn{1}{r}{\small 187.9675} \\
%&  &  & \multicolumn{1}{r}{} & \multicolumn{1}{r}{} & \multicolumn{1}{r}{} &
%\multicolumn{1}{r}{} & \multicolumn{1}{r}{} & \multicolumn{1}{r}{} &
%\multicolumn{1}{r}{} \\
$^{150}${\small Nd} & {\small (a)} &  & \multicolumn{1}{r}{\small 163.8037}
& \multicolumn{1}{r}{} & \multicolumn{1}{r}{\small 55.8968} &
\multicolumn{1}{r}{} & \multicolumn{1}{r}{\small 97.8169} &
\multicolumn{1}{r}{} & \multicolumn{1}{r}{\small 133.6912} \\
& {\small (b)} &  & \multicolumn{1}{r}{\small 130.1364} & \multicolumn{1}{r}{
} & \multicolumn{1}{r}{\small 43.8840} & \multicolumn{1}{r}{} &
\multicolumn{1}{r}{\small 77.3178} & \multicolumn{1}{r}{} &
\multicolumn{1}{r}{\small 105.9993} \\
& {\small (c)} &  & \multicolumn{1}{r}{\small 160.2720} & \multicolumn{1}{r}{
} & \multicolumn{1}{r}{\small 54.6713} & \multicolumn{1}{r}{} &
\multicolumn{1}{r}{\small 95.6942} & \multicolumn{1}{r}{} &
\multicolumn{1}{r}{\small 130.8005} \\
& {\small (d)} &  & \multicolumn{1}{r}{\small 131.9781} & \multicolumn{1}{r}{
} & \multicolumn{1}{r}{\small 44.6741} & \multicolumn{1}{r}{} &
\multicolumn{1}{r}{\small 78.5433} & \multicolumn{1}{r}{} &
\multicolumn{1}{r}{\small 107.5715} \\ \hline\hline
\end{tabular}
\end{table}

\begin{table}[htbp]
\caption{Decomposition of NTMEs for the $\left( \beta ^{-}\beta ^{-}\right)
_{0\nu }$ decay of $^{100}$Mo including finite size effect (F) and SRC (F+S)
for the $PQQ1$ parametrization.}
\label{tab2}
\begin{tabular}{llccccccc}
\hline\hline
NTMEs &  & {\small F} &  &  &  & {\small F+S} &  &  \\ \cline{5-9}
& \multicolumn{1}{c}{} &  &  & {\small SRC1} &  & {\small SRC2} &  & {\small %
SRC3} \\ \hline
$\mathbf{M_{F}}$ & \multicolumn{1}{c}{} & \multicolumn{1}{r}{\small 68.6223} &  &
\multicolumn{1}{r}{\small 35.8191} &  & \multicolumn{1}{r}{\small 54.0101} &
& \multicolumn{1}{r}{\small 64.2516} \\
&&&&&&&&\\
$M_{GT-AA}$ & \multicolumn{1}{c}{} & \multicolumn{1}{r}{\small -370.5960} &
& \multicolumn{1}{r}{\small -144.5650} &  & \multicolumn{1}{r}{\small %
-242.3340} &  & \multicolumn{1}{r}{\small -316.3250} \\
$M_{GT-AP}$ & \multicolumn{1}{c}{} & \multicolumn{1}{r}{\small 174.4640} &
& \multicolumn{1}{r}{\small 43.3631} &  & \multicolumn{1}{r}{\small 93.9737}
&  & \multicolumn{1}{r}{\small 137.5100} \\
$M_{GT-PP}$ & \multicolumn{1}{c}{} & \multicolumn{1}{r}{\small -66.3082} &
& \multicolumn{1}{r}{\small -8.3767} &  & \multicolumn{1}{r}{\small -28.5936}
&  & \multicolumn{1}{r}{\small -48.0727} \\
$M_{GT-MM}$ & \multicolumn{1}{c}{} & \multicolumn{1}{r}{\small -41.7693} &
& \multicolumn{1}{r}{\small 16.3949} &  & \multicolumn{1}{r}{\small 7.6213}
&  & \multicolumn{1}{r}{\small -13.3421} \\
$\mathbf{M_{GT}}$ & \multicolumn{1}{c}{} & \multicolumn{1}{r}{\small -304.2095} &
& \multicolumn{1}{r}{\small -93.1837} &  & \multicolumn{1}{r}{\small %
-169.3326} &  & \multicolumn{1}{r}{\small -240.2298} \\
&&&&&&&&\\
$M_{T-AP}$ & \multicolumn{1}{c}{} & \multicolumn{1}{r}{\small 9.4369} &  &
\multicolumn{1}{r}{\small 9.2393} &  & \multicolumn{1}{r}{\small 10.0332} &
& \multicolumn{1}{r}{\small 10.0610} \\
$M_{T-PP}$ & \multicolumn{1}{c}{} & \multicolumn{1}{r}{\small -3.6622} &  &
\multicolumn{1}{r}{\small -3.5528} &  & \multicolumn{1}{r}{\small -3.9226} &
& \multicolumn{1}{r}{\small -3.9386} \\
$M_{T-MM}$ & \multicolumn{1}{c}{} & \multicolumn{1}{r}{\small 1.2567} &  &
\multicolumn{1}{r}{\small 1.1163} &  & \multicolumn{1}{r}{\small 1.3438} &
& \multicolumn{1}{r}{\small 1.3722} \\
$\mathbf{M_{T}}$ & \multicolumn{1}{c}{} & \multicolumn{1}{r}{\small 7.0314} &  &
\multicolumn{1}{r}{\small 6.8028} &  & \multicolumn{1}{r}{\small 7.4544} &
& \multicolumn{1}{r}{\small 7.4945} \\
&&&&&&&&\\
$\left| M_N^{(0\nu )}\right| $ & \multicolumn{1}{c}{} & \multicolumn{1}{r}%
{\small 365.8004} &  & \multicolumn{1}{r}{\small 122.2000} &  &
\multicolumn{1}{r}{\small 215.8882} &  & \multicolumn{1}{r}{\small 296.9869}
\\ \hline\hline
\end{tabular}

\end{table}

\subsection{Short range correlations and radial evolutions of NTMEs}

In the approximation of finite size of nucleons with dipole form 
factor (F) and finite size plus SRC (F+S), the theoretically calculated twelve NTMEs $%
M_{N}^{\left( 0\nu \right) }$ using the four sets of HFB\ wave functions
generated with $PQQ1$, $PQQHH1$, $PQQ2$ and $PQQHH2$ parameterizations of the
effective two-body interaction and three different parameterizations of
Jastrow type of SRC for $^{94,96}$Zr, $^{98,100}$Mo, $^{104}$Ru, $^{110}$Pd, $%
^{128,130}$Te and $^{150}$Nd isotopes are presented in Table \ref{tab1}. 

To analyze the role of different components of NTME $M_{N}^{\left( 0\nu \right) }$, 
the decomposition 
of the latter into Fermi, different terms of Gamow-Teller and tensor matrix elements of 
$^{100}$Mo are
presented in Table \ref{tab2} for $PQQ1$ parametrization. From the 
inspection of Table \ref{tab2}, the following observations emerge.
 
\begin{enumerate}[(i)]

\item The contribution of conventional Fermi matrix elements $M_{Fh}=M_{F-VV}$ is 
about 20\% to the total matrix element. 

\item The Gamow-Teller matrix element is noticeably modified by the inclusion of the pseudoscalar 
and weak magnetism terms in the hadronic currents. While $M_{GT-PP}$ increases the absolute value of 
 $M_{GT-AA}$, $M_{GT-AP}$ has a significant contribution with opposite sign in all cases.
 The term $M_{GT-MM}$ is smaller than others, and the introduction of short range correlations 
changes its sign.
 
 \item The tensor matrix elements have a very small contribution, smaller than 2\%, to the total 
transition matrix elements.

\item The inclusion of short range correlations changes the nuclear matrix elements significantly, 
whose effects are large for the Gamow-Teller and Fermi matrix elements but small in the case of 
tensor ones.

\item The Miller-Spencer parameterization of the short range correlations, SRC1, cancels out a large 
part of the radial function $H_N$, as shown in Fig. 1. The same cancellation reduces the calculated matrix 
elements to about one third of its original value. The other two parameterizations of the short range 
correlations, namely SRC2 and SRC3, have a sizable effect, which is in all cases much smaller than SRC1.
  
\end {enumerate}

With respect to the point nucleon case. the change in $M_{N}^{\left( 0\nu \right) }$ 
is about 30\%--34\% due to the FNS. With the inclusion of effects due to FNS and SRC, 
the NTMEs change by about 75\%--79\%, 58\%--62\% and 43\%--47\%  
for F+SRC1, F+SRC2 and F+SRC3, respectively.
It is noteworthy that the SRC3 has practically negligible effect on the
finite size case. Further, the maximum variation in $M_{N}^{\left( 0\nu
\right) }$ due to $PQQHH1$, $PQQ2$ and $PQQHH2$ parametrization with respect
to PQQ1 interaction are about 24\%, 18\% and 26\% respectively. 

\begin{figure}[htbp]
\includegraphics [scale=0.38]{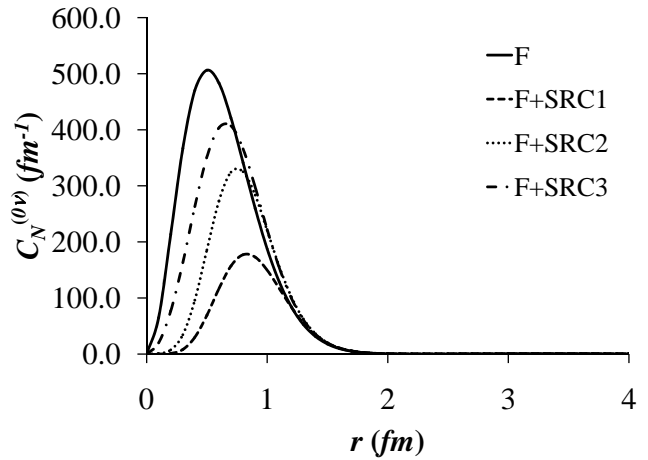} \\
\caption{Radial dependence of $C_{N}^{(0\nu)}(r)$ for the
$\left( \beta^{-}\beta ^{-}\right) _{0\nu }$ decay of $^{100}$Mo isotope.}
\label{fig3}
\end{figure}

\begin{figure*}[htbp]
\begin{tabular}{cc}
\includegraphics [scale=0.37]{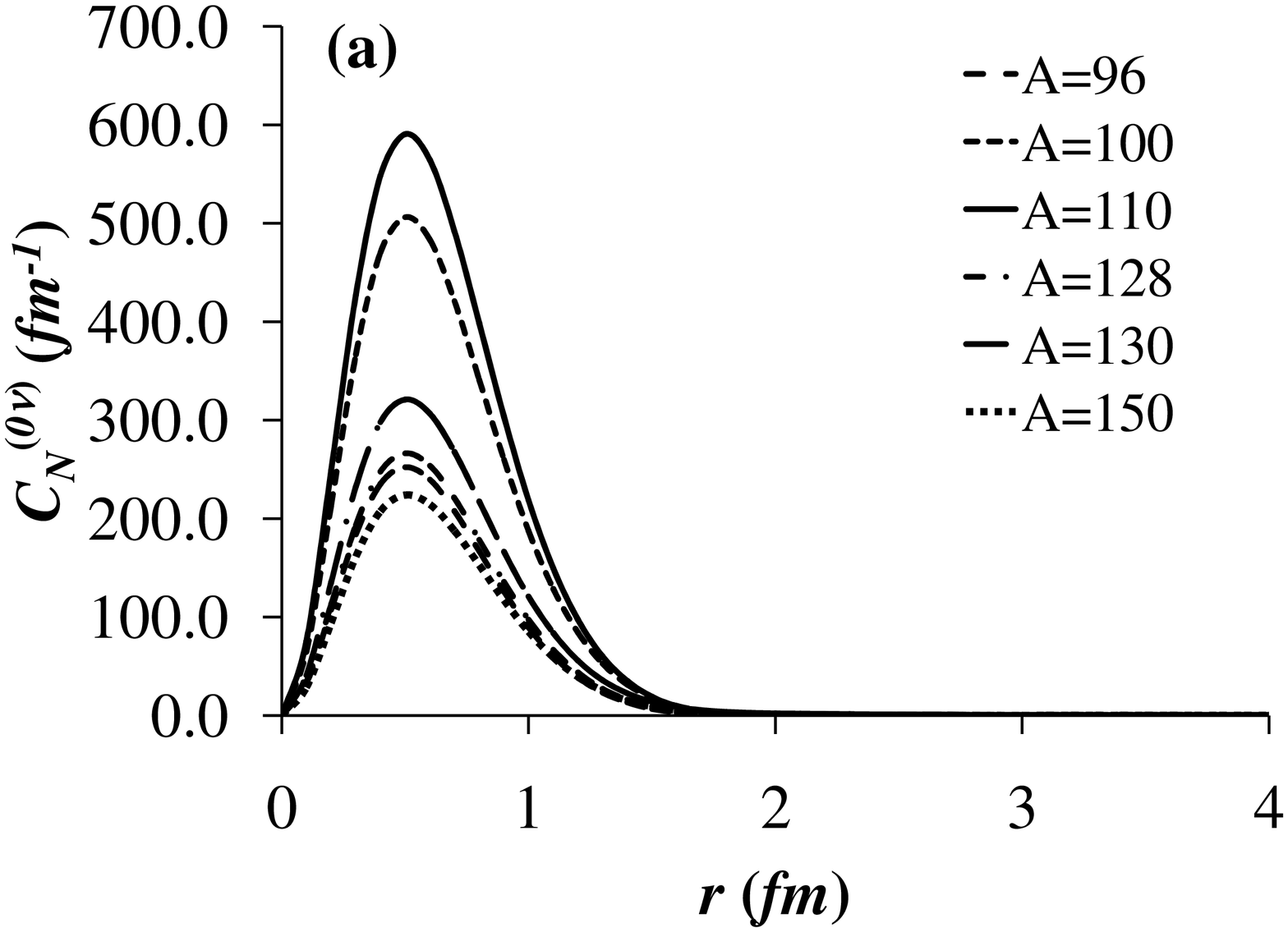} &
\includegraphics [scale=0.37]{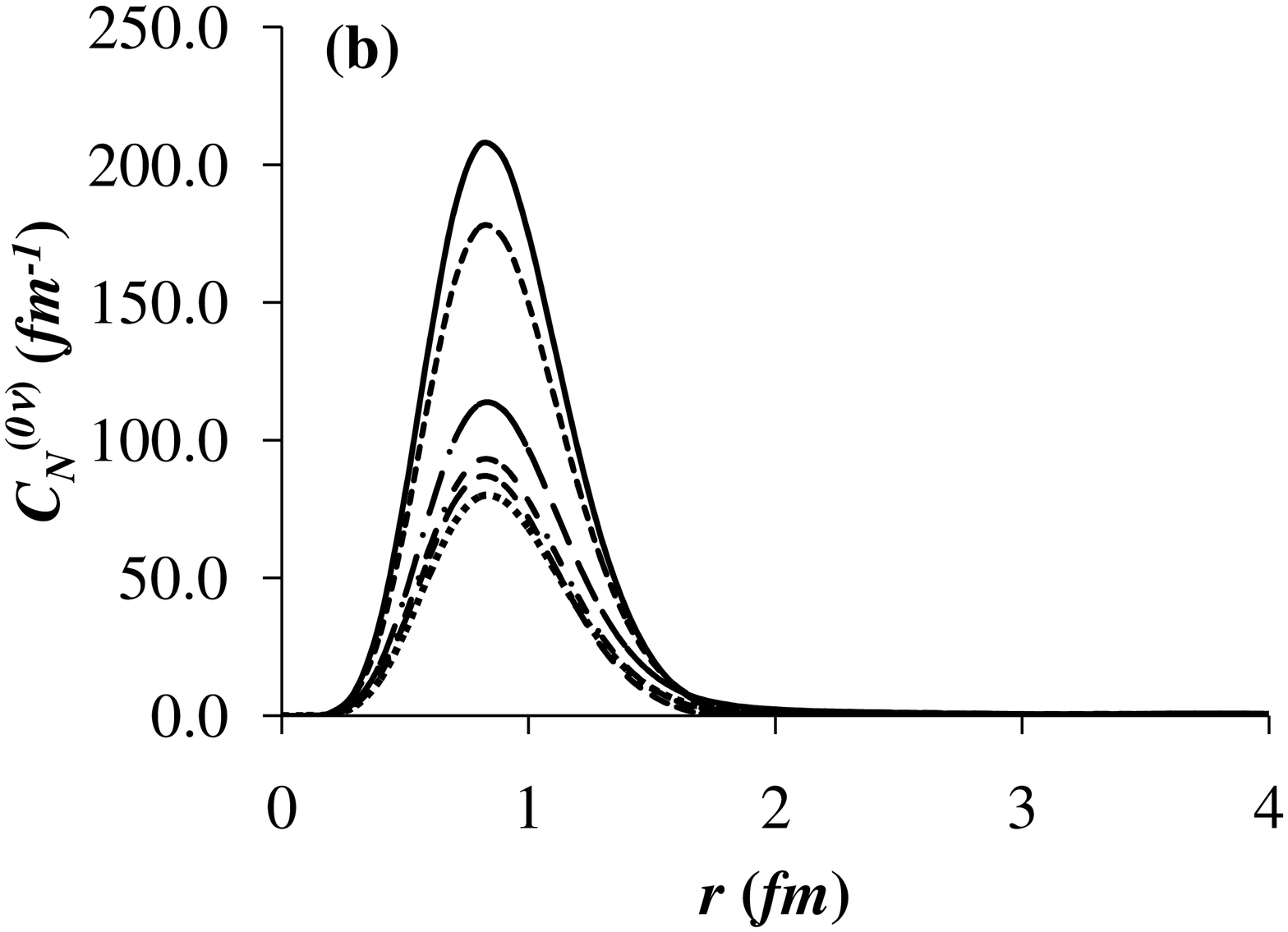} \\
&\\
&\\
\includegraphics [scale=0.37]{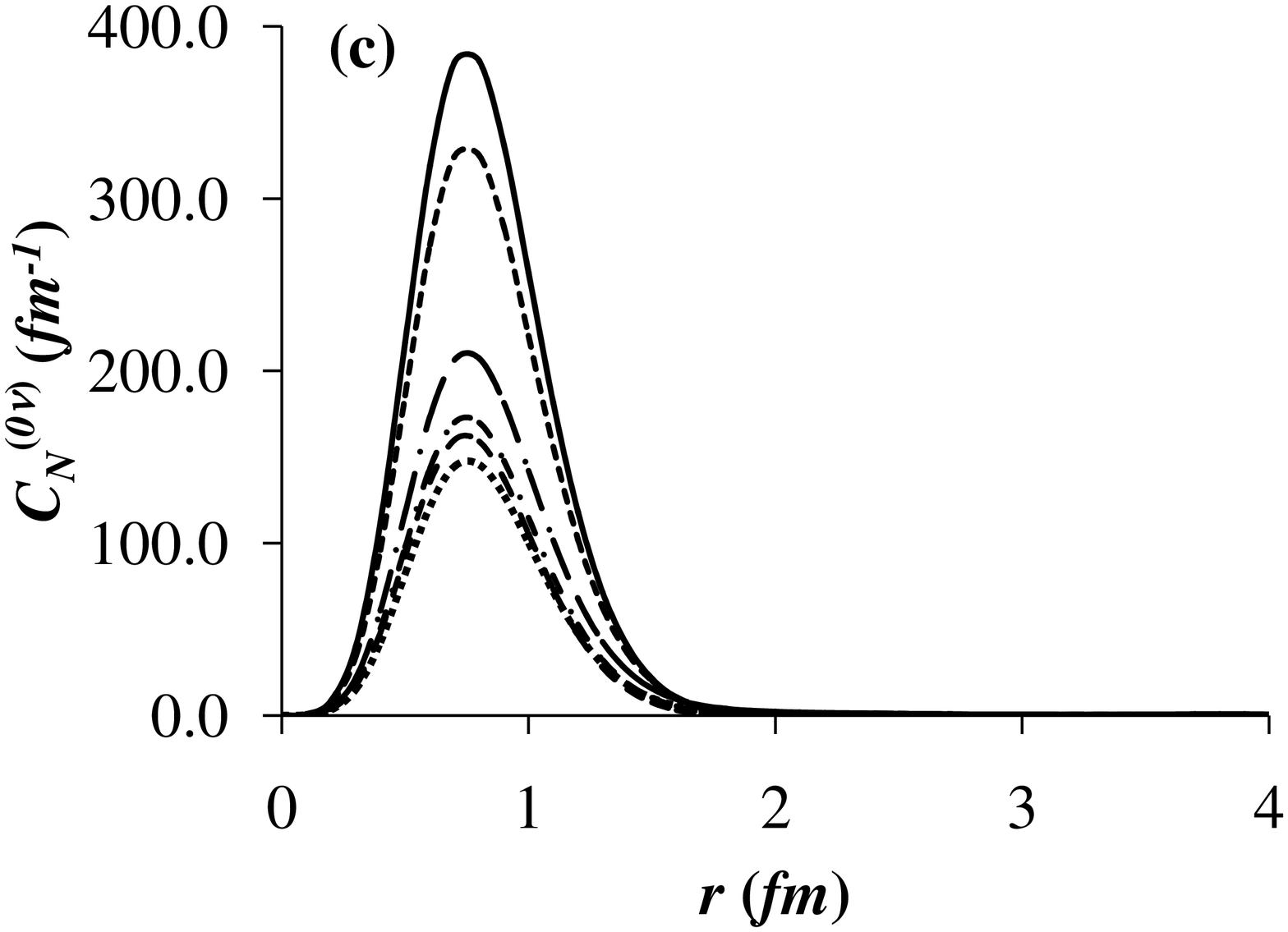} &
\includegraphics [scale=0.37]{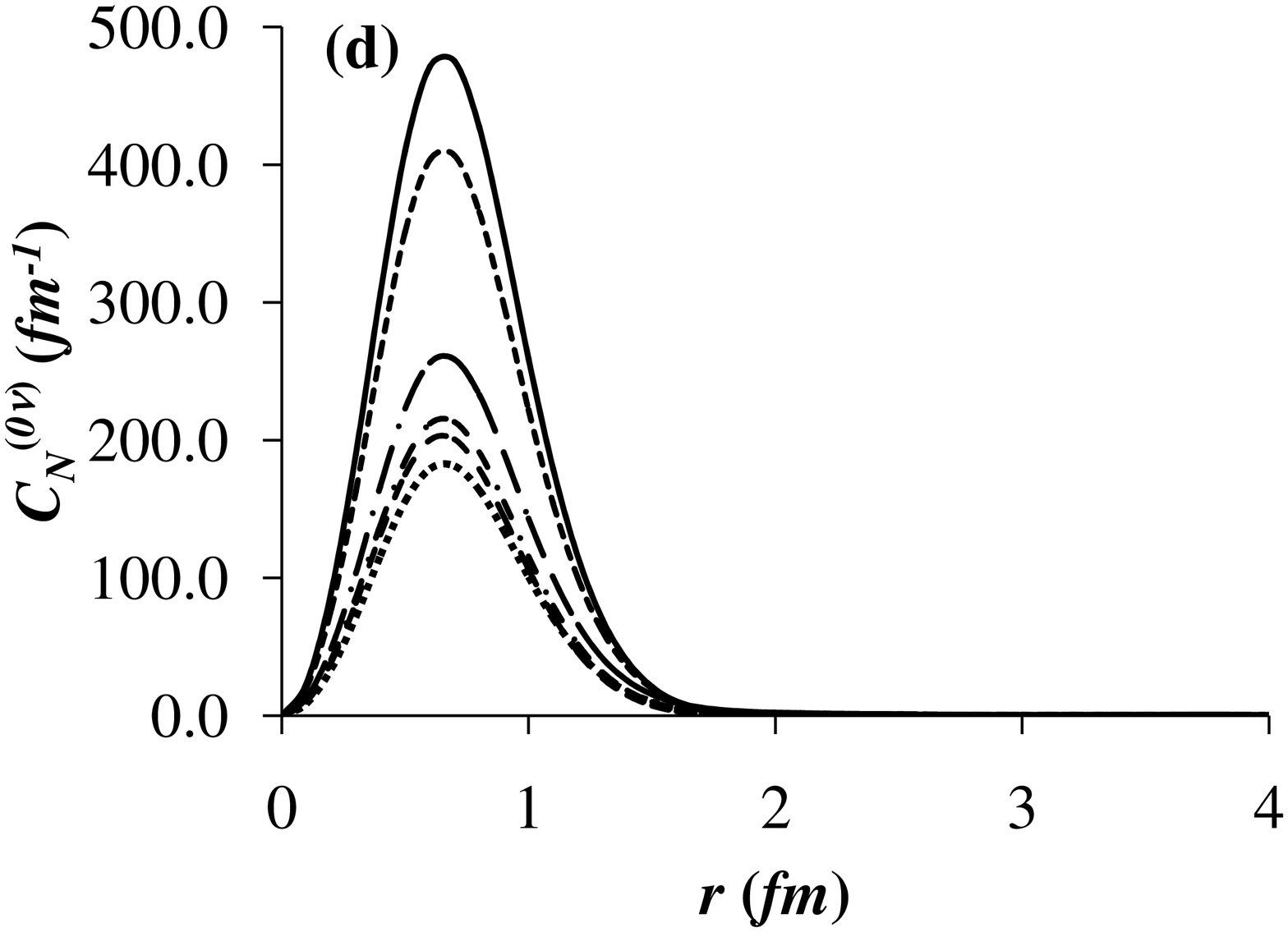} \\
\end{tabular}
\caption{Radial dependence of $C_{N}^{(0\nu)}(r)$ for the
$\left( \beta^{-}\beta ^{-}\right) _{0\nu }$ decay of $^{96}$Zr,
$^{100}$Mo, $^{110}$Pd, $^{128,130}$Te and $^{150}$Nd isotopes.
In this Fig., (a), (b), (c) and (d) correspond to F, F+SRC1, F+SRC2 and
F+SRC3, respectively.}
\label{fig2}
\end{figure*}

In the QRPA \cite{simk08,simk09}, ISM \cite{caur08} and PHFB \cite
{rath09,rath10}, the radial evolution of $M^{\left( 0\nu \right) }$ due to
the exchange of light Majorana neutrino has already been studied. In both
QRPA and ISM calculations, it has been established that the contributions of
decaying pairs coupled to $J=0$ and $J>0$ almost cancel beyond $%
r\approx $3 fm and the magnitude of $C^{\left( 0\nu \right) }$ for all
nuclei undergoing $\left( \beta ^{-}\beta ^{-}\right) _{0\nu }$ decay 
%are the
have their
maximum at about the internucleon distance $r\approx $1 fm. These
observations were also made in the PHFB model \cite{rath09,rath10}.
Similarly, the radial evolution of $M_{N}^{\left( 0\nu \right) }$ can be
studied by defining 
\begin{equation}
M_{N}^{\left( 0\nu \right) }=\int C_{N}^{\left( 0\nu \right) }\left(
r\right) dr
\end{equation}

The radial evolution of $M_{N}^{\left( 0\nu
\right) }$ has been studied for four cases, namely F, F+SRC1, F+SRC2 and
F+SRC3. To make the effects of finite size and SRC more transparent, we plot
them for $^{100}$Mo in Fig. 2. In case of finite sized nucleons, the $%
C_{N}^{\left( 0\nu \right) }$ are peaked at $r\approx $0.5 fm and with
the addition of SRC1 and SRC2, the peak shifts to about 0.8 fm. However, the position of
peak is shifted to 0.7 fm for SRC3. In Fig. 3, 
we plot the radial dependence of $C_{N}^{\left( 0\nu
\right) }$ for six nuclei, namely $^{96}$Zr, $^{100}$Mo, $^{110}$Pd, $%
^{128,130}$Te and $^{150}$Nd and the same observations remain valid. Also, the 
same features in the radial distribution of  $C_{N}^{\left( 0\nu\right) }$ are 
noticed in the cases of $PQQ2$, $PQQHH1$ and $PQQHH2$ parametrizations.

\subsection{Uncertainties in NTMEs}

The uncertainties associated with the NTMEs $M_{N}^{(0\nu )}$ for $\left(
\beta ^{-}\beta ^{-}\right) _{0\nu }$ decay are estimated by preforming a
statistical analysis by using Eqs.(\ref{avrg}) and (\ref{stdv}). 
In Table \ref{tab1}, sets of twelve NTMEs $M_{N}^{(0\nu )}$ of $^{94,96}$Zr, 
$^{98,100}$Mo, $^{110}$Pd, $^{128,130}$Te and $^{150}$Nd isotopes are 
displayed, which are employed to calculate 
the average values $\overline{M}_{N}^{(0\nu )}$ as well as uncertainties 
$\Delta \overline{M}_{N}^{(0\nu)}$ tabulated in Table \ref{tab3} for the bare 
axial vector coupling 
constant $g_{A}=1.254$ and quenched value of $g_{A}=1.0$.

\begin{table}[htbp]
\caption {Average NTMEs $\overline{M}_{N}^{(0\nu )}$ and uncertainties $%
\Delta \overline{M}_{N}^{(0\nu )}$ for the $\left( \beta ^{-}\beta
^{-}\right) _{0\nu }$ decay of $^{94,96}$Zr, $^{98,100}$Mo, $^{104}$Ru, 
$^{110}$Pd, $%
^{128,130}$Te and $^{150}$Nd isotopes. Both bare and quenched values of $%
g_{A}$ are considered. Case I and Case II denote calculations with and
without SRC1, respectively.}
\label{tab3}
\begin{tabular}{clcccccccc}
\hline\hline
$\beta ^{-}\beta ^{-}$ & $g_{A}$ &  & \multicolumn{3}{c}{Case I} & ~~ &
\multicolumn{3}{c}{Case II} \\ \cline{4-6}\cline{8-10}
emitters &  &  & $\overline{M}_{N}^{(0\nu )}$ &~  & $\Delta \overline{M}%
_{N}^{(0\nu )}$ &  & $\overline{M}_{N}^{(0\nu )}$ &~  & $\Delta \overline{M}%
_{N}^{(0\nu )}$ \\ \hline
$^{94}${\small Zr} & {\small 1.254} &  & {\small 126.2146} &  & {\small %
44.9489} &  & {\small 152.8378} &  & {\small 27.1912} \\
& {\small 1.0} &  & {\small 142.9381} &  & {\small 49.1752} &  & {\small %
172.1620} &  & {\small 29.3965} \\
%&  &  &  &  &  &  &  &  &  \\
$^{96}${\small Zr} & {\small 1.254} &  & {\small 100.5313} &  & {\small %
36.8858} &  & {\small 122.5048} &  & {\small 21.9209} \\
& {\small 1.0} &  & {\small 114.4851} &  & {\small 40.3246} &  & {\small %
138.6328} &  & {\small 23.5263} \\
%&  &  &  &  &  &  &  &  &  \\
$^{98}${\small Mo} & {\small 1.254} &  & {\small 202.5006} &  & {\small %
71.6345} &  & {\small 245.3957} &  & {\small 41.8882} \\
& {\small 1.0} &  & {\small 230.1520} &  & {\small 78.3244} &  & {\small %
277.2795} &  & {\small 44.9878} \\
%&  &  &  &  &  &  &  &  &  \\
$^{100}${\small Mo} & {\small 1.254} &  & {\small 206.7533} &  & {\small %
73.0792} &  & {\small 250.1870} &  & {\small 43.7119} \\
& {\small 1.0} &  & {\small 235.0606} &  & {\small 79.9885} &  & {\small %
282.7964} &  & {\small 47.1334} \\
%&  &  &  &  &  &  &  &  &  \\
$^{104}${\small Ru} & {\small 1.254} &  & {\small 150.5572} &  & {\small %
53.9389} &  & {\small 182.7216} &  & {\small 31.9382} \\
& {\small 1.0} &  & {\small 171.8075} &  & {\small 59.0467} &  & {\small %
207.1750} &  & {\small 34.3939} \\
%&  &  &  &  &  &  &  &  &  \\
$^{110}${\small Pd} & {\small 1.254} &  & {\small 231.4743} &  & {\small %
82.4924} &  & {\small 280.5688} &  & {\small 49.1588} \\
& {\small 1.0} &  & {\small 263.4339} &  & {\small 90.3033} &  & {\small %
317.3947} &  & {\small 53.0150} \\
%&  &  &  &  &  &  &  &  &  \\
$^{128}${\small Te} & {\small 1.254} &  & {\small 126.8285} &  & {\small %
46.3381} &  & {\small 153.7370} &  & {\small 29.4676} \\
& {\small 1.0} &  & {\small 143.9772} &  & {\small 50.6942} &  & {\small %
173.5263} &  & {\small 31.8554} \\
%&  &  &  &  &  &  &  &  &  \\
$^{130}${\small Te} & {\small 1.254} &  & {\small 136.3856} &  & {\small %
46.9164} &  & {\small 164.5378} &  & {\small 27.2226} \\
& {\small 1.0} &  & {\small 154.3797} &  & {\small 51.2511} &  & {\small %
185.2849} &  & {\small 29.1907} \\
%&  &  &  &  &  &  &  &  &  \\
$^{150}${\small Nd} & {\small 1.254} &  & {\small 85.5467} &  & {\small %
31.4473} &  & {\small 103.4294} &  & {\small 20.9802} \\
& {\small 1.0} &  & {\small 97.3640} &  & {\small 34.5024} &  & {\small %
117.0160} &  & {\small 22.8729} \\ \hline\hline
\end{tabular}
\end{table}
 
It turns out that in all cases, the uncertainties $\Delta \overline{M}%
^{(0\nu )}$ are about 35\% for $g_{A}=1.254$ and $g_{A}=1.0$. Further, we
estimate the uncertainties for eight NTMEs $\overline{M}_{N}^{(0\nu )}$
calculated using the SRC2, and SRC3 parameterizations and the uncertainties
in NTMEs reduce to about 16\% to 20\% with the exclusion of
Miller-Spenser type of parametrization. In Table \ref{tab4}, 
average NTMEs for case II along with NTMEs calculated in other models have been
presented. It is noteworthy that in the models employed in Refs. 
\cite{tomo91,hirs96,pant94}, effects due to higher order currents have not been included.
We also extract lower
limits on the effective mass of heavy Majorana neutrino $%
\left\langle M_{N}\right\rangle $ from the largest observed limits on
half-lives $T_{1/2}^{0\nu }$ of $\left( \beta ^{-}\beta ^{-}\right) _{0\nu }$
decay. The extracted limits are
$\left\langle M_{N}\right\rangle $ $%
>5.67_{-0.94}^{+0.94}\times 10^{7}$ GeV and $>4.06_{-0.64}^{+0.64}\times $%
{\small 10}$^{7}$ GeV, from the limit on half-life $T_{1/2}^{0\nu }$ $>$%
{\small 3.0}$\times ${\small 10}$^{24}$ yr of $^{130}$Te \cite{arna08} for 
$g_{A}=1.254$ and $g_{A}=1.0$, respectively. 

\begin{table*}[htbp]
\caption {Average NTMEs $\overline{M}_{N}^{\,^{\prime }(0\nu )}\left(
=(g_{A}/1.254)^{2}\overline{M}_{N}^{(0\nu )}\right) $ for the $%
\left( \beta ^{-}\beta ^{-}\right) _{0\nu }$ \ decay of $^{94,96}$Zr, $%
^{98,100}$Mo, $^{110}$Pd, $^{128,130}$Te and $^{150}$Nd isotopes. Both bare
and quenched values of $g_{A}$ are considered. The superscripts $a$ and $b$
denote the Argonne and CD-Bonn potentials.}
\label{tab4}
\begin{tabular}{clccccccccccccccccccc}
\hline\hline
$\beta ^{-}\beta ^{-}$ & $g_{A}$ &  & $\overline{M}_{N}^{\,^{\prime }(0\nu )}$
&  & {\small QRPA} &  & {\small QRPA} &  & {\small QRPA} &  & {\small QRPA}
&  & {\small SRQRPA}$^{a}$ &  & {\small SRQRPA}$^{b}$ &  & $T_{1/2}^{0\nu }(%
\text{ }yr)$ & {\small Ref.} &  & $\left\langle m_{N}\right\rangle ${\small %
(GeV)} \\
emitters &  &  &  &  & \cite{tomo91} &  & \cite{hirs96} &  & \cite{pant94} &
& \cite{pant96} &  & \cite{simk10} &  & \cite{simk10} &  &  &  &  &  \\
\hline
$^{94}${\small Zr} & {\small 1.254} &  & {\small 152.84}$\pm ${\small 27.19}
&  &  &  &  &  &  &  &  &  &  &  &  &  & {\small 1.9}$\times ${\small 10}$%
^{19}$ & \cite{arno99} &  & \multicolumn{1}{l}{{\small 2.57}$%
_{-0.46}^{+0.46}\times ${\small 10}$^{4}$} \\
& {\small 1.0} &  & {\small 109.48}$\pm ${\small 18.69} &  &  &  &  &  &  &
&  &  &  &  &  &  &  &  &  & \multicolumn{1}{l}{{\small 1.84}$%
_{-0.31}^{+0.31}\times ${\small 10}$^{4}$} \\
$^{96}${\small Zr} & {\small 1.254} &  & {\small 122.50}$\pm ${\small 21.92}
&  &  &  &  &  &  &  & {\small 99.062} &  &  &  &  &  & {\small 9.2}$\times $%
{\small 10}$^{21}$ & \cite{argy10} &  & \multicolumn{1}{l}{{\small 2.68}$%
_{-0.48}^{+0.48}\times ${\small 10}$^{6}$} \\
& {\small 1.0} &  & {\small 88.16}$\pm ${\small 14.96} &  &  &  &  &  &  &
&  &  &  &  &  &  &  &  &  & \multicolumn{1}{l}{{\small 1.93}$%
_{-0.33}^{+0.33}\times ${\small 10}$^{6}$} \\
$^{98}${\small Mo} & {\small 1.254} &  & {\small 245.40}$\pm ${\small 41.89}
&  &  &  &  &  &  &  &  &  &  &  &  &  & {\small 1.0}$\times ${\small 10}$%
^{14}$ & \cite{frem52} &  & \multicolumn{1}{l}{{\small 9.70}$_{-1.70}^{+1.70}
$} \\
& {\small 1.0} &  & {\small 176.33}$\pm ${\small 28.61} &  &  &  &  &  &  &
&  &  &  &  &  &  &  &  &  & \multicolumn{1}{l}{{\small 6.97}$%
_{-1.13}^{+1.13}$} \\
$^{100}${\small Mo} & {\small 1.254} &  & {\small 250.19}$\pm ${\small 43.71}
&  & {\small 155.960} &  & {\small 333.0} &  & {\small 56.914} &  & {\small %
76.752} &  & {\small 259.8} &  & {\small 404.3} &  & {\small 4.6}$\times $%
{\small 10}$^{23}$ & \cite{arno05} &  & \multicolumn{1}{l}{{\small 3.43}$%
_{-0.60}^{+0.60}\times ${\small 10}$^{7}$} \\
& {\small 1.0} &  & {\small 179.84}$\pm ${\small 29.97} &  &  &  &  &  &  &
&  &  & {\small 191.8} &  & {\small 310.5} &  &  &  &  & \multicolumn{1}{l}{%
{\small 2.47}$_{-0.41}^{+0.41}\times ${\small 10}$^{7}$} \\
$^{110}${\small Pd} & {\small 1.254} &  & {\small 280.57}$\pm ${\small 49.16}
&  &  &  &  &  &  &  &  &  &  &  &  &  & {\small 6.0}$\times ${\small 10}$%
^{17}$ & \cite{wint52} &  & \multicolumn{1}{l}{{\small 2.43}$%
_{-0.43}^{+0.43}\times ${\small 10}$^{4}$} \\
& {\small 1.0} &  & {\small 201.84}$\pm ${\small 33.71} &  &  &  &  &  &  &
&  &  &  &  &  &  &  &  &  & \multicolumn{1}{l}{{\small 1.75}$%
_{-0.29}^{+0.29}\times ${\small 10}$^{4}$} \\
$^{128}${\small Te} & {\small 1.254} &  & {\small 153.74}$\pm ${\small 29.47}
&  & {\small 122.669} &  & {\small 303.0} &  &  &  & {\small 101.233} &  &
&  &  &  & {\small 1.1}$\times ${\small 10}$^{23}$ & \cite{arna03} &  &
\multicolumn{1}{l}{{\small 2.06}$_{-0.39}^{+0.39}\times ${\small 10}$^{6}$}
\\
& {\small 1.0} &  & {\small 110.35}$\pm ${\small 20.26} &  &  &  &  &  &  &
&  &  &  &  &  &  &  &  &  & \multicolumn{1}{l}{{\small 1.48}$%
_{-0.27}^{+0.27}\times ${\small 10}$^{6}$} \\
$^{130}${\small Te} & {\small 1.254} &  & {\small 164.54}$\pm ${\small 27.22}
&  & {\small 108.158} &  & {\small 267.0} &  &  &  & {\small 92.661} &  &
{\small 239.7} &  & {\small 384.5} &  & {\small 3.0}$\times ${\small 10}$%
^{24}$ & \cite{arna08} &  & \multicolumn{1}{l}{{\small 5.67}$%
_{-0.94}^{+0.94}\times ${\small 10}$^{7}$} \\
& {\small 1.0} &  & {\small 117.83}$\pm ${\small 18.56} &  &  &  &  &  &  &
&  &  & {\small 176.5} &  & {\small 293.8} &  &  &  &  & \multicolumn{1}{l}{%
{\small 4.06}$_{-0.64}^{+0.64}\times ${\small 10}$^{7}$} \\
$^{150}${\small Nd} & {\small 1.254} &  & {\small 103.43}$\pm ${\small 20.98}
&  & {\small 153.085} &  & {\small 422.0} &  &  &  &  &  &  &  &  &  &
{\small 1.8}$\times ${\small 10}$^{22}$ & \cite{argy09} &  &
\multicolumn{1}{l}{{\small 5.99}$_{-1.21}^{+1.21}\times ${\small 10}$^{6}$}
\\
& {\small 1.0} &  & {\small 74.41}$\pm ${\small 14.55} &  &  &  &  &  &  &
&  &  &  &  &  &  &  &  &  & \multicolumn{1}{l}{{\small 4.31}$%
_{-0.84}^{+0.84}\times ${\small 10}$^{6}$} \\ \hline\hline
\end{tabular}

\end{table*}

\section{CONCLUSIONS}
 
We have employed the PHFB model, with four different parameterizations of pairing plus
multipole effective two body interaction, to generate sets of four HFB intrinsic wave 
functions, which reasonably
reproduced the observed spectroscopic properties, namely the yrast spectra,
reduced $B(E2$:$0^{+}\rightarrow 2^{+})$ transition probabilities, static
quadrupole moments $Q(2^{+})$ and $g$-factors $g(2^{+})$ of participating
nuclei in $\left( \beta ^{-}\beta ^{-}\right) _{2\nu }$ \ decay, as well as their $%
M_{2\nu }$ \cite{chan05,sing07}. Considering three different parameterizations 
of Jastrow type of
SRC, sets of twelve NTMEs $M_{N}^{\left( 0\nu \right) }$ for the study $%
\left( \beta ^{-}\beta ^{-}\right) _{0\nu }$ \ decay of $^{94,96}$Zr, $%
^{98,100}$Mo, $^{104}$Ru, $^{110}$Pd, $^{128,130}$Te and $^{150}$Nd isotopes
in the heavy Majorana neutrino mass mechanism have been calculated.

The study of effects due to finite size of nucleons and SRC reveal that in
the case of heavy Majorana neutrino exchange, the NTMEs change by about
30\%--34\% due to finite size of nucleons and the SRC1, SRC2 and SRC3 change
them by 75\%--79\%, 58\%--62\% and 43\%--47\%, respectively. Further, it has
been noticed through the study of radial evolution of NTMEs that the FNS and
SRC play a more crucial role in the heavy than in the light Majorana
neutrino exchange mechanism. 

Finally, a statistical analysis has been performed by employing the sets of
twelve NTMEs $M_{N}^{\left( 0\nu \right) }$ to estimate the uncertainties
for $g_{A}=1.254$ and $g_{A}=1.0$. It turns out that the uncertainties are
about 35\% for all the considered nuclei. Exclusion of Miller-Spenser
parametrization of Jastrow type of SRC, reduces the maximum uncertainties to
a value smaller than 20\%.
The best extracted limit on
the effective heavy Majorana neutrino mass $\left\langle M_{N}\right\rangle $
from the available limits on experimental half-lives $T_{1/2}^{0\nu }$ using
average NTMEs $\overline{M}_{N}^{(0\nu )}$ calculated in the PHFB model is
$>5.67_{-0.94}^{+0.94}\times 10^{7}$ GeV and $>4.06_{-0.64}^{+0.64}\times $%
{\small 10}$^{7}$ GeV for $^{130}$Te isotope.

\begin{acknowledgments}
This work is partially supported by the Department of Science and
Technology (DST), India vide sanction No. SR/S2/HEP-13/2006, DST-RFBR
Collaboration via grant no. RUSP-935, Consejo Nacional de Ciencia y 
Tecnolog\'{i}a (Conacyt)-M\'{e}xico, European Union-Mexico Science and 
Technology International Cooperation Fund (FONCICYT) Project 94142, 
and Direcci\'{o}n General de Asuntos del Personal Acad\'{e}mico, 
Universidad Nacional Aut\'{o}noma de M\'{e}xico (DGAPA-UNAM) Project IN102109-3.
\end{acknowledgments}

\end{document}